\newif\ifjournal  % JOURNAL SUBMISSION?

\ifjournal
  \documentclass[conference]{IEEEtran}
  \IEEEoverridecommandlockouts
\else
  \documentclass[twoside,leqno,twocolumn]{article}
  \usepackage{orcidlink}
\fi
\usepackage{cite}
\usepackage{amsmath,amssymb,amsfonts}
\usepackage{algorithmic}
\usepackage{graphicx}
\usepackage{textcomp}
\usepackage{xcolor}
\usepackage{tabularx}
\usepackage{makecell}
\usepackage[table]{xcolor}
\usepackage{booktabs}
\usepackage{multirow}
\usepackage{hyperref}
\usepackage{orcidlink}
\usepackage{pgfplots}
\pgfplotsset{compat=1.18}

\usepackage{fancyhdr}
 
\usepgfplotslibrary{colormaps}

\def\BibTeX{{\rm B\kern-.05em{\sc i\kern-.025em b}\kern-.08em
    T\kern-.1667em\lower.7ex\hbox{E}\kern-.125emX}}

\colorlet{coolingrevision}{blue}

\hypersetup{hidelinks}
\newif\ifblind
\begin{document}

\makeatletter
\def\blfootnote{\xdef\@thefnmark{}\@footnotetext}
\makeatother

\makeatletter
    \newcommand{\linebreakand}{%
      \end{@IEEEauthorhalign}
      \hfill\mbox{}\par
      \mbox{}\hfill\begin{@IEEEauthorhalign}
    }
    \makeatother

%Energy-aware Performance Characterization of Kokkos Kernels Across Precision Levels
\title{Energy-Aware Performance Characterization of Kokkos Kernels Across Precision Levels
%\\
}

\ifjournal
\unless\ifblind
\author{\IEEEauthorblockN{Maria Patrou \orcidlink{0000-0003-3975-4638}}
\IEEEauthorblockA{
\textit{Oak Ridge National Laboratory}\\
Oak Ridge, TN, USA \\
patroum@ornl.gov}
\and
\IEEEauthorblockN{Andrey Prokopenko \orcidlink{0000-0003-3616-5504}}
\IEEEauthorblockA{
\textit{Oak Ridge National Laboratory}\\
Oak Ridge, TN, USA \\
prokopenkoav@ornl.gov}
\and
\IEEEauthorblockN{Aditya Kashi \orcidlink{0000-0003-2589-3792}}
\IEEEauthorblockA{
\textit{Oak Ridge National Laboratory}\\
Oak Ridge, TN, USA \\
kashia@ornl.gov}
% \and
\linebreakand
\IEEEauthorblockN{Oscar Hernandez \orcidlink{0000-0002-5380-6951}} 
\IEEEauthorblockA{
\textit{Oak Ridge National Laboratory}\\
Oak Ridge, TN, USA \\
oscar@ornl.gov}
\and
\IEEEauthorblockN{Jean-Luc Fattebert \orcidlink{0000-0002-1455-6160}}
\IEEEauthorblockA{
\textit{Oak Ridge National Laboratory}\\
Oak Ridge, TN, USA \\
fattebertj@ornl.gov}
}
\fi
\else
  \author{
    M.~Patrou\thanks{Oak Ridge National Laboratory}\enskip\orcidlink{0000-0003-3975-4638},
    A.~Prokopenko\footnotemark[1]\enskip\orcidlink{0000-0003-3616-5504},
    A.~Kashi\footnotemark[1]\enskip\orcidlink{0000-0003-2589-3792},
    O.~Hernandez\footnotemark[1]\enskip\orcidlink{0000-0002-5380-6951},
    J.-L.~Fattebert\footnotemark[1]\enskip\orcidlink{0000-0002-1455-6160}
  }
\fi

\maketitle
\begin{abstract}
Kokkos is a performance-portable programming model used in GPU-accelerated HPC applications comprising kernels with diverse computational and memory behaviors. We characterize kernels from two applications to evaluate how FP32, FP64, and mixed FP32/FP64 affect arithmetic intensity, performance throughput, and GPU energy consumption. We introduce an Energy-Aware Roofline Model that incorporates energy into performance analysis and a Thermal-Aware Inter-Kernel Energy Model that visualizes energy consumption alongside duration, GPU temperature, and execution order.
We study both applications with: (1) fine-grained, application-level instrumentation through the Kokkos Performance Tools PAPI Connector and (2) coarse-grained PAPI instrumentation applied to a dedicated-kernel benchmark using vendor-specific power management components. 
Our analysis shows that precision effects vary by application, kernel, and workload. Mixed FP32/FP64 kernels generally achieve arithmetic intensity comparable to FP32, with modestly lower throughput and, in most cases, a modest increase in energy consumption. FP64 kernels generally exhibit the highest energy consumption among the evaluated precision configurations. 
%Similar sampled temperatures can accompany substantially different energy costs. 
%Temperature reflects execution history and thermal interactions among regions with different energy consumption, whereas runtime and energy more directly identify optimization priorities.
\end{abstract}

\ifjournal
\fi

\unless\ifblind

\ifjournal
  %copyright
  \thispagestyle{fancy}
  \lhead{}
  \rhead{}
  \chead{}
  \lfoot{\footnotesize{
  SC26 Workshops, November 15-20, 2026, Chicago, Illinois, USA
  \newline 979-8-3195-1221-5/26/\$31.00 \copyright 2026 IEEE}}
  \rfoot{}
  \cfoot{}
  \renewcommand{\headrulewidth}{0pt}
  \renewcommand{\footrulewidth}{0pt}
\fi

\blfootnote {%
% ORNL disclaimer
This manuscript has been authored by UT-Battelle, LLC, under contract
DE-AC05-00OR22725 with the U.S. Department of Energy. The United States
Government retains and the publisher, by accepting the article for
publication, acknowledges that the United States Government retains a
nonexclusive, paid-up, irrevocable, world-wide license to publish or reproduce
the published form of this manuscript, or allow others to do so, for United
States Government purposes. The DOE will provide public access to these results
in accordance with the DOE Public Access Plan
(http://energy.gov/downloads/doe-public-access-plan).
}
\fi

\section{Introduction}
\label{sec:introduction}

Energy consumption has become a first-order concern in high-performance computing (HPC) systems and data centers, on par with raw performance. Modern supercomputers face strict power budgets (e.g., a 30 MW cap for exascale systems), making energy efficiency essential. The Green500 list ranks supercomputers by ``FLOPS per Watt'', highlighting the push for maximizing computational work per unit energy. Consequently, HPC application developers must optimize not only runtime performance but also energy usage.

Scientific applications often consist of multiple computational kernels or phases, each with different characteristics (compute-intensive, memory-intensive, network-intensive etc.). Traditionally, performance models such as the Roofline model~\cite{williams2009roofline} help developers identify performance bottlenecks by relating attainable performance throughput (in flop/s) to a kernel’s arithmetic intensity (flop/byte). By plotting these quantities and comparing them against hardware ceilings, namely peak compute capability and peak memory bandwidth, the model reveals whether a kernel is compute-bound or memory-bound, and how far it is from the achievable limits of the architecture. However, data points in the Roofline model alone do not quantify the energy required to complete a kernel. 
For a complete performance analysis, developers need to compare throughput with completion time and energy usage, then locate costly regions within application execution.

We present two complementary performance visualization models for this task on graphics processing units (GPUs). The Energy-Aware Roofline Model connects arithmetic intensity and achieved throughput to measured GPU energy. The Thermal-Aware Inter-Kernel Energy Model compares region duration and energy while recording their temperatures and execution order. 

This paper makes the following contributions:

\begin{itemize}
    \item Energy-Focused Models:
    We combine energy-aware Roofline analysis with a view of region/kernel energy, duration, execution order, and temperature. Power reference lines distinguish average power from accumulated energy. Together, the models identify costly computation and describe its execution context, including regions without floating-point operations. We evaluate this characterization capability.
    \item Multi-Application Precision Analysis:
    We analyze kernels from two applications (LOBPCG and ArborX) at various floating-point precisions in data and operations (FP32 vs FP64) to expose the impact of precision on time-to-solution, energy, power, and temperature. 
    At a time when scientists are looking at how to incorporate low precision and mixed precision algorithms in their computations \cite{mxp_review_2025}, our approach provides more insight into the impact of these techniques.
    Eigensolvers of the Locally Optimal Block Preconditioned Conjugate Gradient (LOBPCG) type are at the core of Density Functional Theory calculations that use a large fraction of HPC resources at supercomputing centers \cite{levine2026openmolecules2025omol25}.
    \item Energy-Aware Instrumentation Methodology Comparison:
    We compare the region-level method with Kokkos-tools and PAPI used in ArborX with the dedicated-kernel benchmark used in the LOBPCG eigensolver. We provide advantages and limitations when using both techniques to collect and analyze the data.
\end{itemize}

Together, these contributions address three research questions:
\begin{description}
\item[RQ1:] How does incorporating energy consumption into Roofline analysis provide a broader multi-objective view of individual GPU kernel behavior?
\item[RQ2:] What does the Thermal-Aware Inter-Kernel Energy Model reveal about region/kernel energy costs and temperature during application execution?
\item[RQ3:] How do numerical precision and workload affect kernel-level performance, energy consumption, and thermal behavior across Kokkos-based GPU applications?
\end{description}

\section{Background}
\label{sec:background}

\subsection{Roofline Model}
The roofline model was introduced by Williams et al.~\cite{williams2009roofline} to visualize the limiting factor for the performance of a computational kernel.
The concept of arithmetic intensity (or operational intensity) of a kernel is key to this method, typically defined as the number of floating-point operations (flop) performed by the kernel per byte of data loaded from memory.
The roofline method involves plotting the flop throughput of a kernel against its arithmetic intensity, superimposed on the machine's \emph{roofline}.

If the kernel's arithmetic intensity (and the $x$-axis of the plot) is $\alpha$, the processor hardware's peak memory bandwidth is $m$ and peak flop throughput is $f$, then the roofline is $\mathrm{min}\{m\alpha, f\}$.

Traditionally, performance optimization attempts to move the kernel as close to the roofline as possible, while also increasing the arithmetic intensity as much as possible.
For a fixed problem size, this is expected to help achieve faster time-to-solution.
While theoretical values of arithmetic intensity can sometimes be used for these models based on the knowledge of the specific operations involved in an algorithm, those can also be measured and provide additional insights\cite{Ofenbeck2014}.

\subsection{Performance Monitoring Methodologies}
Runtime data collection combines hardware-assisted monitoring with software instrumentation. External devices, off-chip analyzers, and on-chip performance counters can observe system-wide events with little application overhead, but they may be costly, platform-specific, or limited to summarized data. Software instrumentation complements these sources by delimiting code regions and controlling when counters are read~\cite{pierce1995instrumentation}. Software-only methods range from high-overhead code emulation to instrumentation inserted through microcode, operating-system traps, executables, object files, or source code. Direct code instrumentation is widely applicable because it preserves program logic while trading modest runtime overhead for portable, region-level attribution~\cite{pierce1995instrumentation}.

\subsection{Performance Application Programming Interface (PAPI) - NVIDIA Management Library (NVML)}
The Performance Application Programming Interface (PAPI) is an interface that provides performance counter monitoring capabilities as a library and a tool for performance analysis~\cite{jagode2025advancements}. It offers two APIs for selecting and accessing hardware counters: a low-level (e.g., PAPI\_read) and a high-level (HL) API (e.g., PAPI\_hl\_region\_begin, PAPI\_hl\_region\_end) for instrumented code regions~\cite{winkler2020redesigning}.
The advantage of using the high-level API lies in its ease of use, as it requires less configuration while internally invoking the low-level one; however, both APIs can be used simultaneously, if needed~\cite{papi_hl_git}.
Through these interfaces, PAPI enables users to access performance and power counters across diverse software technologies and hardware architectures, including NVIDIA GPUs through the integration of CUDA Profiling Tools Interface (CUPTI)’s API, the Nsight Perf SDK, and the NVIDIA Management Library (NVML). Compute performance data such as number of single-precision and double-precision floating-point operations, and energy/thermal data—such as power consumption, fan speed, and temperature are exposed via the CUDA and NVML components, respectively~\cite{jagode2025advancements}. The NVML API provides methods for different purposes, including Device Queries, to access information against each (GPU) device. PAPI integrates queries, such as {\it nvmlDeviceGetPowerUsage} and {\it nvmlDeviceGetTemperature}, enabling users to select the hardware counters and devices to monitor by defining the corresponding PAPI\_EVENTS. The NVML API also offers additional device queries, such as {\it nvmlDeviceGetTotalEnergyConsumption} that retrieves the total energy consumption for a selected GPU device in millijoules (mJ) since the driver was last reloaded~\cite{nvml_ref}.

\subsection{Kokkos}

Kokkos~\cite{kokkos} is a library allowing abstraction of common parallel
execution patterns, such as parallel loops, reductions, and scans (prefix
sums), from the underlying hardware. Kokkos provides an abstraction for a
multi-dimensional data structure called \texttt{View}. The layout of this data
structure depends on which memory the data resides in (host or device).

The Kokkos Core library allows applications to instrument their runtime through
the Kokkos profiling interface. The interface allows an application to dynamically
load helper tools to be executed in certain places, such as beginning and end
of any parallel region, memory allocations, and data transfers. This allows
developers to collect kernel-level performance and debugging data without
requiring changes to the build. A collection of such tools is
available in a separate Kokkos Tools project, including many tools for
debugging and profiling~\cite{osti_1656942}.

We note two tools of particular interest to our paper. The \emph{space-time
stack} profiling tool provides a nested kernel/region representation, including
the number of calls, execution time, total application percentage, and memory
information. The \emph{PAPI connector} uses an external PAPI installation's
high-level (HL) API to access PAPI hardware counters that are specified via the
\textit{PAPI\_EVENTS} environment variable. Upon application finalization, a
JSON file is generated reporting the measured events for each parallel
construct and profiling section~\cite{kokkos_papi_connector}.

\section{Related Work}
\label{sec:related}

\subsection{Roofline and Energy Models}

Choi et al.~\cite{choi2013energyroofline} proposed a new roofline model of energy, considering the differences that arise when one tries to optimize for energy rather than time as the constraining resource.
They point out that the traditional roofline model looks at a kernel in relation to the \emph{time balance} $B_\tau$ of the machine, which is the ratio of the peak flops the machine can perform \emph{per unit time} to the peak bytes it can fetch from memory per unit time.
They look at the \emph{energy balance} $B_\epsilon$ of the machine, which is the ratio of the peak flops \emph{per unit energy consumed} to the peak bytes fetched from memory per unit energy.
They show that when $B_\tau \geq B_\epsilon$, optimizing for time will lead towards optimal energy efficiency as well. Otherwise, optimizing for the fastest execution may increase energy efficiency, but further optimizations may be possible that yield even better energy efficiency without speeding up the execution further.
Their methodology involves dedicated hardware for physically measuring the current and voltage to the GPU, CPU, etc. Their studies involve the NVIDIA GTX 580 GPU of the Fermi architecture as well as an Intel workstation CPU of the Nehalem generation.

Ding et al.~\cite{9059264} proposed an Instruction Roofline Model for GPU-accelerated applications that captures instruction throughput, thread predication and memory access efficiency across the memory hierarchies by counting instructions and memory transactions. They introduced Instruction Intensity (warp instructions per memory transaction), measured performance in billions of instructions per second (GIPS) scaled to billions of transactions per second (GTXN/s), and defined memory walls to represent and categorize memory access patterns. They evaluated their model on NVIDIA V100 GPUs with five proxy applications.

When it comes to energy efficiency, one can use hand-derived analytical quantities, such as the theoretical number of flops or bytes, as a proxy for energy consumption. However, this depends on application-specific knowledge and may be limited in real-world accuracy. For example, Anzt et al. \cite{anzt2019adaptive} reported on the modeled energy consumption of a mixed precision preconditioner that they proposed for use in sparse linear solvers. They assume that the workload is always limited by main memory bandwidth and use the total number of bits of data as a proxy for the relative energy consumption of the solver.

Ghane et al.~\cite{ghane2018power} extended the Roofline model to visualize the relationship between energy efficiency and power consumption for GPUs. They developed computational and memory roofline models with axes representing energy per flop and per byte transferred, against power (W). They incorporated optimization techniques such as fused multiply-add (FMA), instruction level parallelism (ILP), thread and block scaling, and strided memory accesses as ceilings, and collected energy data using their Phoenix library. They compared single- and double-precision kernels on two NVIDIA GPU architectures. 

Ilic et al.~\cite{ilic2013cache} built the Cache-Aware Roofline Model (CARM) that incorporates complex memory hierarchies in a single plot by accounting for all memory operations, including accesses to different cache levels, with bandwidth defined relative to the processor core and dependent on the accessed memory level for CPUs. They further extended the CARM model in~\cite{ilic2016beyond} by developing PowerCARM, a set of cache-aware power, energy, and energy-efficiency models that visualize the power consumption, energy usage, and energy efficiency of CPU cores, off-core components (uncore), and the overall chip package; the plots use arithmetic intensity on the x-axis and power, energy, or flop per unit energy on the y-axis for different hardware domains, while Intel RAPL (Running Average Power Limit) was used for measurement collection and validation on Intel 3770K (Ivy Bridge) processor.

Energy estimates obtained by integrating built-in GPU power-sensor traces face a different limitation: the samples can be coarse-grained (e.g. millisecond scale) or filtered. Yang et al.~\cite{10793163} show that, on NVIDIA A100 and H100 GPUs, \textit{nvidia-smi} reports power as a 25~ms average sampled every 100~ms. As a result, power samples can miss short bursts of activity and underestimate energy consumption.

We therefore obtain per-kernel energy from \textit{nvmlDeviceGetTotalEnergyConsumption} through \textit{PAPI\_EVENTS} rather than integrating sampled power. We add measured energy to the Roofline view and introduce a visualization model that presents region-level energy and duration relative to power reference lines while retaining Kokkos execution order and temperature.

\subsection{Profiling Tools}

Dietrich et al.~\cite{dietrich2021enabling} implemented a Score-P adapter for the Kokkos profiling interface to enable performance analysis of Kokkos applications. The adapter was built as a shared library, loaded via the KOKKOS\_PROFILE\_LIBRARY environment variable. The authors collected execution traces and visualized them with Vampir. This enabled analysis of Kokkos region behavior and its correlation with the underlying CUDA backend, while it could be extended in the future to include additional information for mapping results back to the source code.

PowerPack~\cite{4906989} is an integrated hardware and software framework for power and energy profiling in distributed systems. It combines sensors, digital meters, APIs, and analysis tools to correlate system and application activities with power consumption at both component and code-segment granularity. Experimental data are collected remotely via a local intranet (out-of-band measurements), while user-level APIs synchronize application execution with the power profiling process.

Chen et al. \cite{chen2025nekbone} utilize a new tool called Verificarlo and develop a promising methodology to instrument an application to study changes in numerical behavior in response to precision changes. In addition, they use cache-aware roofline modeling to determine possible improvements in time-to-solution from using lower precision. In contrast to this work, while their focus on accuracy is notable, they focus exclusively on CPUs, use Cray PM counters, and do not show a unified picture of energy and performance throughput (flop/s).

Our profiling methodology relies on PAPI through two approaches: Kokkos Performance Tools for application-level instrumentation and a dedicated-kernel benchmark.

\subsection{Application-Aware Thermal Management}
\label{sec:thermal_related}

Application-aware thermal management couples a temperature objective to the performance cost of changing execution. Sarood and Kal\'{e}~\cite{sarood2012cooldown} exploit differences in frequency sensitivity among application regions to apply Dynamic Voltage and Frequency Scaling (DVFS) under a core-temperature constraint. Menon et al.~\cite{menon2013thermal} combine temperature control with automated load balancing to reduce the timing penalty caused by unequal processor speeds. Their runtime accounts for the cost of monitoring and redistribution when deciding when to intervene. These studies motivate distinguishing application phases, since phases can respond differently to frequency changes. Choosing a control action requires considering its expected performance impact and monitoring or redistribution overhead, in addition to the measured temperature.

In our application-level GPU measurements, named regions identify the kernels being measured, and execution order records their sequence. The Energy-Aware Roofline Model relates computational behavior to GPU energy, while the Thermal-Aware Inter-Kernel Energy Model locates region costs in execution order and records their thermal context. This representation supports examining how temperature varies across application regions with different computational behavior.

\section{Methodology}
\label{sec:methodology}

\subsection{Energy-focused Instrumentation Techniques}

We first use the Kokkos Tools space-time stack to identify the kernels that dominate execution from the call counts and timings in its Top-Down Tree. We then collect per-kernel FP32 and FP64 operation counts and memory traffic from CUDA counters, and temperature from NVIDIA Management Library (NVML) counters via PAPI.

%low level papi apl: read
%Average (mean): 6.3066 ms
%Sample standard deviation: 0.0155 ms
%run 11 times on hudson no exlusive access, Andrey's  benchmark

For energy, we read the cumulative GPU energy counter exposed by NVML at region boundaries and compute $E_\mathrm{region}=E_\mathrm{end}-E_\mathrm{start}$ through PAPI HL. The \textit{nvmlDeviceGetTotalEnergyConsumption} API reports the energy consumed since the driver was last reloaded, in millijoules, on supported Volta or
newer GPUs~\cite{nvml_ref}. These differences directly measure each region's energy consumption without integrating separately sampled power measurements over time. They remain subject to the counter's update behavior. We collect the counter through PAPI, which we extended to expose this NVML metric.
We also modified the Kokkos Tools PAPI connector to report kernel-region execution order for extracting inter-kernel relationships.

We use the connector through the PAPI high-level (HL) API to collect per-region/per-kernel measurements. In our setup, instrumentation incurs 6.31~ms of overhead per measured region. We also account for the update frequency (100~ms validated through measurements) of NVML power counters on modern GPUs; shorter intervals may fail to capture new counter updates and can degrade measurement quality ~\cite{mlenergy_profiling_2026,mlenergy_tutorial_2025,nvidia_kernel_modules}. 
We measured the energy counter update interval across 100 transitions under both idle and active GPU workloads. The median interval ranged from 100.01 to 100.05 ms across the two H100 NVIDIA GPUs, corresponding to an update frequency of approximately 10 Hz.
We restrict direct measurements to regions lasting at least 101~ms and discard unchanged energy-counter readings. The reported 6.31~ms instrumentation cost corresponds to approximately 6.25\% of that duration in our experimental setup. This ratio characterizes timing overhead. For shorter regions, we use a dedicated repeated-execution benchmark.

Application-level measurements retain region order, whereas the dedicated benchmark characterizes kernels under repeated execution. Its PAPI low-level API reads counters around a loop that repeatedly invokes each kernel, amortizing boundary-read overhead. We evaluate FP32, FP32 storage with FP64 computation (FP32/FP64), and FP64. This approach provides average per-kernel metrics but does not capture the effects of region ordering and interactions during a complete application run.

Temperature and energy have different measurement semantics. NVML provides an instantaneous temperature reading, while we compute energy consumption as the difference between cumulative energy-counter readings. As with other measurements, we report the mean and sample standard deviation of the recorded temperature readings across runs. The energy measurements cover the GPU device.

\subsection{Energy-Aware Modeling}
We characterize kernels on multiple aspects to help us understand performance behavior and guide optimization efforts. 
We present two energy-aware performance visualization models:
    
\subsubsection*{Energy-Aware Roofline Model} 
The model
extends the traditional Roofline Model by incorporating energy as a third dimension. It enables kernels to be characterized by their arithmetic intensity, achieved performance throughput, and GPU energy consumption. By combining these dimensions, kernels can be categorized according to their execution behavior (memory- vs. compute-bound) while simultaneously comparing their energy consumption (high vs. low energy consumers). This unified view provides deeper insight into kernel behavior, helps identify high-energy kernels, and prioritize optimization opportunities. 

\subsubsection*{Thermal-Aware Inter-Kernel Energy Model} 
The model illustrated in Figure~\ref{fig:energy_model} compares measured region energy against duration and uses color to encode GPU temperature; temperature supplies descriptive execution context. 
% subject to the limitations above. 
The model includes regions without floating-point operations, which cannot be positioned on a floating-point Roofline. For duration $t$ and energy $E$, the ratio $E/t$ gives average GPU power. Lines at idle power and Thermal Design Power (TDP), each multiplied by $t$, provide power references rather than guaranteed energy bounds or definitions of efficiency. A region’s energy represents its accumulated consumption, while its proximity to a reference line indicates its average power, enabling the model to visually classify kernels by their energy and power profiles (e.g., low- versus high-power and low- versus high-energy) and thermal characteristics (e.g., cold versus hot).

Arrows record execution order among regions. Solid lines indicate consecutive kernel executions with no intervening kernels, whereas dashed lines indicate that other kernels are executed in between. The execution order can reveal potential measurement dependencies among kernels and highlight similarities that may extend across neighboring kernels.
They do not represent heat transfer, a continuous timeline, or demonstrated causal influence. The pink area indicates measurements excluded from direct energy comparisons under our measurement policy described above to ensure reliable analysis.

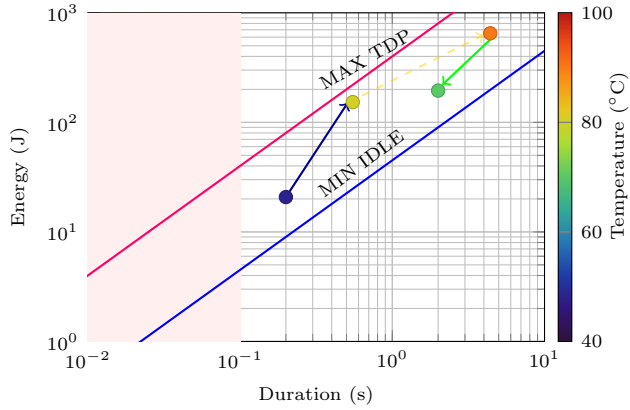
\begin{figure}[htbp]
\centering
\begin{tikzpicture}
\begin{axis}[
    width=0.9\columnwidth,
    height=0.7\columnwidth,
    xlabel={Duration (s)},
    ylabel={Energy (J)},
    ymode=log,
    xmode=log,
    xmin=0.01,
    xmax=10.0,
    ymin=1,
    ymax=1e3,
    grid=both,
    % major grid style={gray!20},
    % minor grid style={gray!10},
    ticklabel style={font=\scriptsize},
    label style={font=\scriptsize},
    point meta min=40,
    point meta max=100,
    % ---- Turbo colormap (compact version) ----
    colormap={turbo}{
        rgb255(0cm)=(48,18,59);
        rgb255(0.1cm)=(50,21,125);
        rgb255(0.2cm)=(31,60,170);
        rgb255(0.3cm)=(38,120,180);
        rgb255(0.4cm)=(50,170,150);
        rgb255(0.5cm)=(90,200,100);
        rgb255(0.6cm)=(170,220,50);
        rgb255(0.7cm)=(230,200,30);
        rgb255(0.8cm)=(250,140,20);
        rgb255(0.9cm)=(240,80,20);
        rgb255(1cm)=(180,20,20);
    },
    colorbar,
    colorbar style={
        ylabel={Temperature ($^\circ$C)},
        yticklabel pos=right,
        ylabel style={
            rotate=0,
            anchor=south,
            font=\scriptsize,
            xshift=8pt,
            yshift=-6pt
        },
        title style={font=\scriptsize},
        ticklabel style={font=\scriptsize},
        width=0.18cm,
        at={(1.03,0.5)},
        anchor=west
    },
]

% Light pink shaded region (Duration <= 0.101)
\fill[pink!25]
    (axis cs:0.001,1e-2)
    rectangle
    (axis cs:0.101,1e3);
     
% MAX TDP line: E = 400 * x
\addplot[
    red!60!magenta,
    thick,
    domain=0.001:10.0
] {400 * x};

\node[rotate=30, anchor=west, font=\scriptsize]
    at (axis cs:0.3, 400*0.5) {MAX TDP};

% MIN IDLE line: E = 45 * x
\addplot[
    blue,
    thick,
    domain=0.001:10.0
] {45 * x};

\node[rotate=36, anchor=west, font=\scriptsize]
    at (axis cs:0.3, 45*0.45) {MIN IDLE};

% Solid arrows between nodes
\draw[->, thick, blue!55!black]
    (axis cs:0.2,20.8) --
    (axis cs:0.51,151.0);

\draw[<-, thick, green]
    (axis cs:2.13,220.0) --
    (axis cs:4.4,570.0);

% dashed arrows between nodes
% \draw[->,thick,dashed,green,opacity=0.5]
%     (axis cs:2,195.0) --
%     (axis cs:0.58,155.0);

\draw[->,thick, dashed, yellow!70!orange,opacity=0.5]
    (axis cs:0.51,151.0) --
    (axis cs:4,630.0);

% Four points with colormap (temperature)
\addplot[
    scatter,
    only marks,
    mark=*,
    mark size=2.5pt,
    scatter src=explicit
]
coordinates {
    (0.2,20.8)  [48]
    (0.55,153.0)  [80]
    (2,195.0)  [70]
    (4.4,650.0) [90]
};

\end{axis}
\end{tikzpicture}
\caption{Schematic of the Thermal-Aware Inter-Kernel Energy Model, separating accumulated energy (J), duration (s), and temperature ($^\circ$C, color). Points are illustrative, not measurements. Lines provide idle-power and TDP references. The pink area marks the excluded duration range. Solid and dashed arrows indicate consecutive and nonconsecutive execution order.}
\label{fig:energy_model}
\end{figure}

\section{Evaluation}
\label{sec:evaluation}

We examine the main computational kernels of two real-world scientific applications using region/kernel level profiling techniques and performance visualization models to provide actionable performance insights.

\subsection{Applications}
\label{se:app}
We study two scientific applications.

\subsubsection*{Block Iterative Eigensolvers}
Iterative eigensolvers are typically used to solve large-scale eigenvalue problems, for computing a few extreme eigenvalues and their associated eigenvectors of a large sparse matrix.
Block solvers are a category of solvers for which multiple eigenvalue/eigenvector pairs are computed simultaneously.
They can be numerically more efficient since they rely principally on higher arithmetic intensity kernels.
Here we focus on the Locally Optimal Block Preconditioned Conjugate Gradient (LOBPCG)~\cite{lobpcg}, an efficient solver in particular for applications where a good preconditioner is available (see~\cite{Knyazev2007} for a reference implementation).
The main data structures are {\it tall-and-skinny} matrices of size $M\times N$ used to store $N$ trial eigenvectors, residuals or other quantities with $M$ components ($M\gg N$).

The main computational kernels in LOBPCG are a sparse matrix-vector multiplication (\emph{spmv}), two flavors of matrix-matrix multiplications ($N \times M$ multiplied by $M \times N$ referred to as \emph{gemm}, and $M \times N$ multiplied by a triangular part of $N \times N$ referred to as \emph{trmm}), an application of a preconditioner (\emph{prec}), and an orthogonalization of a set of $N$ vectors (\emph{Cholesky-QR}).
Other operations are important from an algorithmic point of view, but contribute very little to total computational cost, and are ignored here.
For this study, we use our own Kokkos implementations because vendor libraries do not support all of the mixed-precision variants considered. The resulting \textit{gemm} implementation is memory-bound due to optimizations targeting the ``tall-and-skinny'' matrices used in LOBPCG, as confirmed by the roofline analysis in the next section.

\subsubsection*{ArborX}

ArborX~\cite{arborx2.0} is a performance-portable geometric search library
based on Kokkos. In our experiments, we focus on the DBSCAN clustering
algorithm~\cite{dbscan} implemented in ArborX~\cite{arborx_dbscan}. The
algorithm splits a set of given multi-dimensional points into disjoint sets,
called clusters, grouping together high-density regions of the data. The choice
of this algorithm is motivated by the very different nature of the algorithm
compared to the LOBPCG application. Here, there are only a few computational
kernels; each executed once, with the main task concerning a parallel tree
(bounding volume hierarchy) traversal. We run the algorithm using two datasets
coming from a cosmology application~\cite{arborx_hacc}. 

The main computational kernels are the search index construction
(\textit{BVH::BVH}), computation of the number of neighbors within a given
radius (\textit{DBSCAN::clusters::num\_neigh}), assignment of points to the
clusters by merging the neighbors (\textit{DBSCAN::clusters::query}), and
filtering out small clusters (\textit{DBSCAN::postprocess}). Both
\textit{DBSCAN::clusters::num\_neigh} and \textit{DBSCAN::clusters::query} are
memory-dominant kernels requiring random access during a binary tree traversal.

Both applications can be executed in FP32 and FP64 modes, in which computation and data storage use the same level of precision.
The LOBPCG kernels examined here can run in a mixed precision mode, in which two precisions are used: the precision of the data, and the precision of computations.
For example, \textit{LOBPCG::CholeskyQR FP32/FP64} means that the vectors are stored as float and the computations are done in double.
Depending on precision choices, LOBPCG may become unstable, in particular if FP32 is used everywhere.
In this paper, we analyze the LOBPCG kernels regardless of the stability and accuracy effects on the whole algorithm, which is outside the scope of this paper.
Recently, a mixed-precision implementation of this algorithm was proposed \cite{kressner2023}.

\subsection{Environment Setup}
We run our benchmark suite on a compute node equipped with an AMD EPYC 9454 (Genoa) CPU and two NVIDIA H100 NVL GPUs. 
We use Kokkos \textit{v5.0.0}, 
modified Kokkos Tools % with short commit id \textit{6493964} 
and PAPI,
%at commit \textit{fb07bc4ff}, 
\unless\ifblind which contains our contributions~\cite{pr_525,pr_529}\fi.
%and also includes the fixes from Pull Requests \textit{\#564} and \textit{\#601}
Kokkos Tools and PAPI were run with their default settings.

We run each benchmark seven times and calculate averages and standard deviations (the average values of the various metrics are used in the plots). PAPI events are grouped into different categories: e.g. FP32 operations, FP64 operations, DRAM operations, energy-thermal operations, and retrieved separately.
The ArborX collection script groups runs by precision and workload, with a one-second pause between repetitions. This procedure does not establish a common starting temperature. 
Energy and instantaneous temperature are collected in the same group and attributed to the corresponding regions of the measurement.

For ArborX (commit \textit{c831cd3}), we ran the DBSCAN benchmark with the DBSCAN parameter \emph{minPts} set to 10 (i.e., each point needs to have 10 neighbors within $\varepsilon$ radius to be considered a core point) to collect per-kernel performance information.
For FP32 and FP64 cases, we set the corresponding \textit{precision} flag affecting the data precision used during the hierarchy construction as well as the hierarchy traversal. We consider two problem sizes from~\cite{arborx_hacc}: 37M (using $\varepsilon = 0.042$) and 497M (using $\varepsilon = 0.014$).

For LOBPCG, we evaluated two problem sizes, referred to as ``small'' and ``large'' with sizes 
% -n 16000000 -m 160
% -n 4000000 -m 40
$M=4,000,000$, $N=40$ and $M=16,000,000$, $N=160$ respectively.
The matrices correspond to discretized Hamiltonians in Density Functional Theory \cite{cances2023}, with $M$ being the size of the discretization mesh and $N$ the number of electronic wave functions of interest in these chemistry applications (and also the number of eigenpairs computed).
Only the diagonal elements of the matrices are actually stored, while the rest of it is a finite difference stencil used directly to compute a matrix-vector operation. 

Finally, for ArborX, we use the modified Kokkos Tools PAPI connector to record and present the regions in the order observed by the tool, representing the kernel execution order.

\subsection{Application Kernel Analysis} 
Initial statistical analysis of small and large workloads across both precision levels showed that LOBPCG contains approximately 40 short-lived kernels mostly called at every iteration, and a typical total number of iterations ranging from 50 to 100. The average invocation count per kernel/region and run ranged from one to approximately 2,220, depending on the workload. Among these, \textit{BLAS::dot}, \textit{BLAS::gemm} with any parameters, \textit{BLAS::trmm}, \textit{BLAS::spmv}, \textit{Preconditioner::apply}, and \textit{CholeskyQR} accounted for the majority of the execution time.

Similarly, analysis of small and large workloads across both precision levels showed that ArborX contains 25 kernels, most of which are invoked only once during application execution. Application execution times ranged from 0.5 to 18.5~s, depending on the workload and precision. Kernel execution times exhibited low standard deviation across runs. Among these, \textit{DBSCAN::clusters::num\_neigh}, \textit{DBSCAN::clusters::query}, \textit{DBSCAN::postprocess}, and \textit{BVH::BVH} accounted for the majority of the execution time within the main computational region, \textit{DBSCAN::total}.

Overall, we selected distinct kernels contributing more than $5\%$ of the total computational time in each application for further analysis.

\subsection{Application Kernel Characterization}
Figure~\ref{fig:lobpcg_roof_lg} displays the Energy-Aware DRAM Roofline model of LOBPCG kernels for the large workload.
Regarding the kernels \textit{BLAS::dot}, \textit{BLAS::gemm} and \textit{CholeskyQR} the mixed-precision FP32/FP64 implementations exhibit arithmetic intensities very similar with those of the FP32 implementations. 
Except for \textit{BLAS::dot}, the FP32/FP64 kernels consume slightly more energy than their FP32 counterparts up to 11.52\% (\textit{CholeskyQR}), while achieving slightly lower performance throughput, with reductions of up to 10.76\% in every case. 
In contrast, the FP64 implementations exhibit substantially lower arithmetic intensity and performance throughput than both the FP32 and FP32/FP64 implementations while consuming considerably more energy. Specifically, arithmetic intensity decreases by 50\% (\textit{BLAS::dot}) and by as much as 64.2\% (\textit{Preconditioner::apply}) and performance throughput decreases by 35.2\% (\textit{BLAS::dot}) and by as much as 50.3\% (\textit{BLAS::gemm}). 
Energy consumption increases by 60.3\% for \textit{BLAS::dot} and by approximately a factor of two for \textit{BLAS::gemm}.

Regarding the kernels' execution characteristics, the kernels lie approximately parallel to the GPU memory-bandwidth roof in the memory-bound region. Their clustering along this trend suggests that the application's dominant kernels exhibit a similar relationship between arithmetic intensity and achieved performance, reflecting a common memory-bandwidth-limited execution pattern with this workload. 
This memory-bound behavior also explains the observed precision trends, since memory accesses are dominated by FP32 data even in FP32/FP64 kernels, while FP64 kernels require larger memory transfers for FP64 arrays.
%No strong relationship was observed between energy consumption and either arithmetic intensity or performance throughput.

%our medium load, 
%kernels dot/4000000/float/float: 8270600.0, 5581.142857142857
%kernels dot/4000000/float/double: 8270600.0, 18185.714285714286
%kernels dot/4000000/double/double: 8270600.0, 63960000.0
For the small workload, performance throughput and arithmetic intensity remain similar between most FP32 and FP32/FP64 kernels, consistent with the large-workload results for most kernels. However, \textit{BLAS::dot} exhibits the largest differences in arithmetic intensity among all three precisions due to substantial differences in the amount of DRAM data transferred (\textit{cuda:::dram\_\_bytes}), ranging from 5,581 bytes for FP32 to 63,960,000 bytes for FP64. This difference can be explained by the fact that the working vectors can reside in L2 cache for the small workload, thereby reducing the need to fetch data from DRAM; the L2 cache of the NVIDIA H100 GPU being 50 MB. Energy consumption is substantially higher for FP64 than for the other two precisions, reaching up to 2.86× higher for \textit{BLAS::dot}, while arithmetic intensity and performance throughput for the rest of the kernels decrease for FP64 by up to 84.2\% and 44.9\%, respectively (\textit{BLAS::gemm}).

% Numbers of the plot
% AI	Perf	Energy
% LOBPCG::BLAS::dot FP32	2.52E-01	5.61E+11	2.39E-02
% LOBPCG::BLAS::dot FP32 FP64	2.52E-01	5.36E+11	2.36E-02
% LOBPCG::BLAS::dot FP64	1.26E-01	3.48E+11	3.84E-02
% LOBPCG::BLAS::gemm FP32	4.40E-01	9.09E+11	3.64E+02
% LOBPCG::BLAS::gemm FP32 FP64	4.39E-01	8.80E+11	3.76E+02
% LOBPCG::BLAS::gemm FP64	1.75E-01	4.51E+11	7.28E+02
% LOBPCG::BLAS::trmm FP32	9.12E-01	1.84E+12	9.05E+01
% LOBPCG::BLAS::trmm FP64	4.56E-01	9.88E+11	1.68E+02
% LOBPCG::BLAS::spmv FP32	1.93E+00	2.37E+12	1.07E+01
% LOBPCG::BLAS::spmv FP64	7.49E-01	1.40E+12	1.79E+01
% LOBPCG::Preconditioner::apply FP32	1.93E+00	2.37E+12	1.08E+01
% LOBPCG::Preconditioner::apply FP64	6.88E-01	1.37E+12	1.81E+01
% LOBPCG::CholeskyQR FP32	6.17E-01	1.20E+12	2.77E+02
% LOBPCG::CholeskyQR FP32 FP64	6.09E-01	1.07E+12	3.09E+02
% LOBPCG::CholeskyQR FP64	2.62E-01	6.24E+11	5.31E+02
\begin{figure}[htbp]
    \centering
    \includegraphics[trim=20 22 34 42,clip, width=\linewidth]{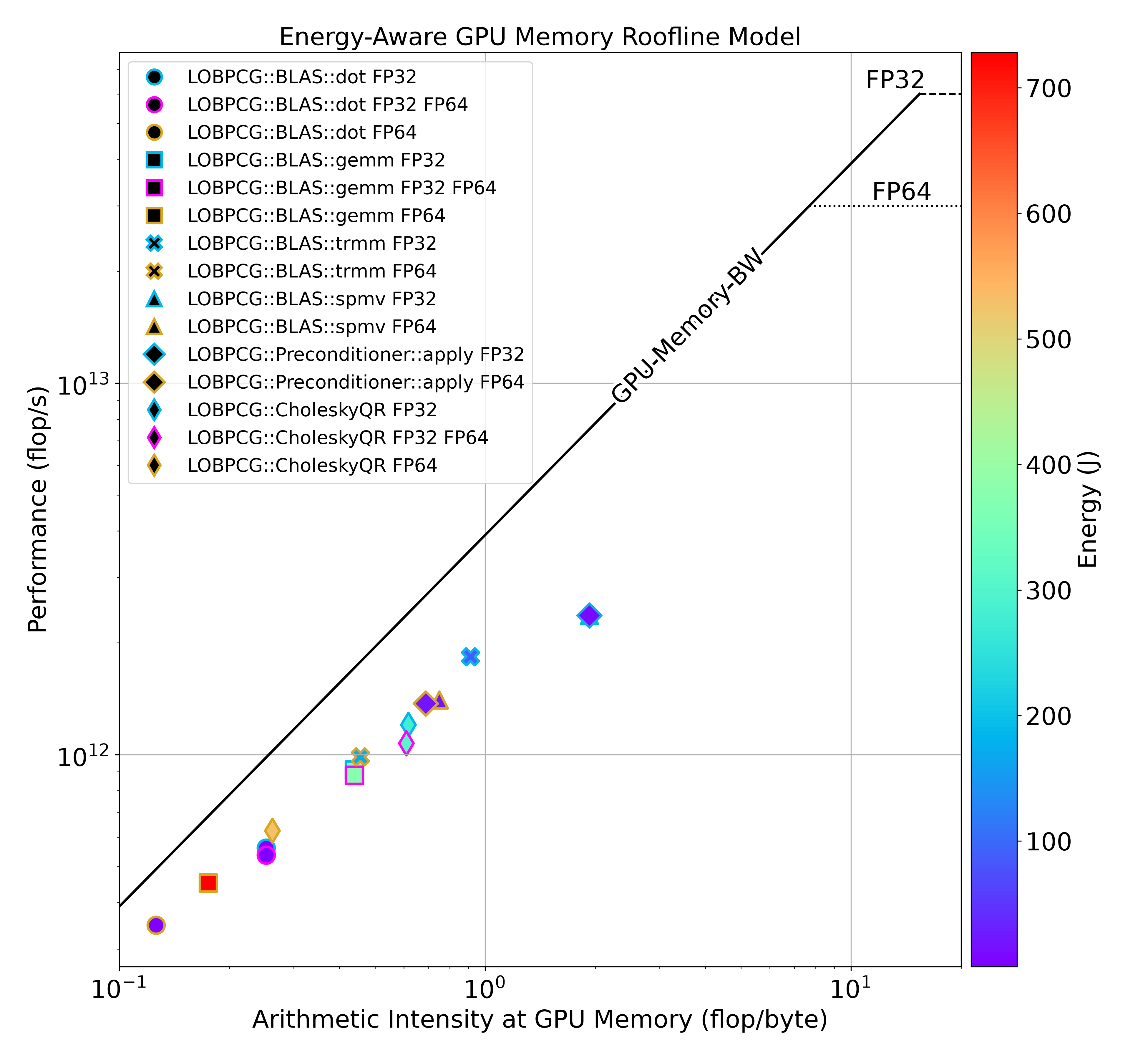}
    \caption{LOBPCG - Energy-Aware Roofline Model Large Workload. Outline color denotes precision (blue for FP32, magenta for FP32/FP64 and yellow for FP64); fill color denotes energy according to the color bar; marker shape denotes the kernel/region.}
    \label{fig:lobpcg_roof_lg}
\end{figure}

% ArborX::BVH::BVH Float, 0.02713432794879243,13867567547.006323, 57.94285714285714
% ArborX::BVH::BVH Double, 0.01898546514317869,11790303792.192944, 84.596
% ArborX::DBSCAN::clusters::num_neigh Float, 7.805106940845209,1383621684318.3164, 160.2467142857143
% ArborX::DBSCAN::clusters::num_neigh Double, 9.348063456489525,1793834916832.8174, 223.55585714285712
% ArborX::DBSCAN::clusters::query Float, 16.5434070203112,840852116001.5424, 705.8514285714285
% ArborX::DBSCAN::clusters::query Double, 11.35356424861162,830754694385.6139, 1012.6365714285715
\begin{figure}[htbp]
    \centering
    \includegraphics[trim=20 22 30 48,clip, width=\linewidth]{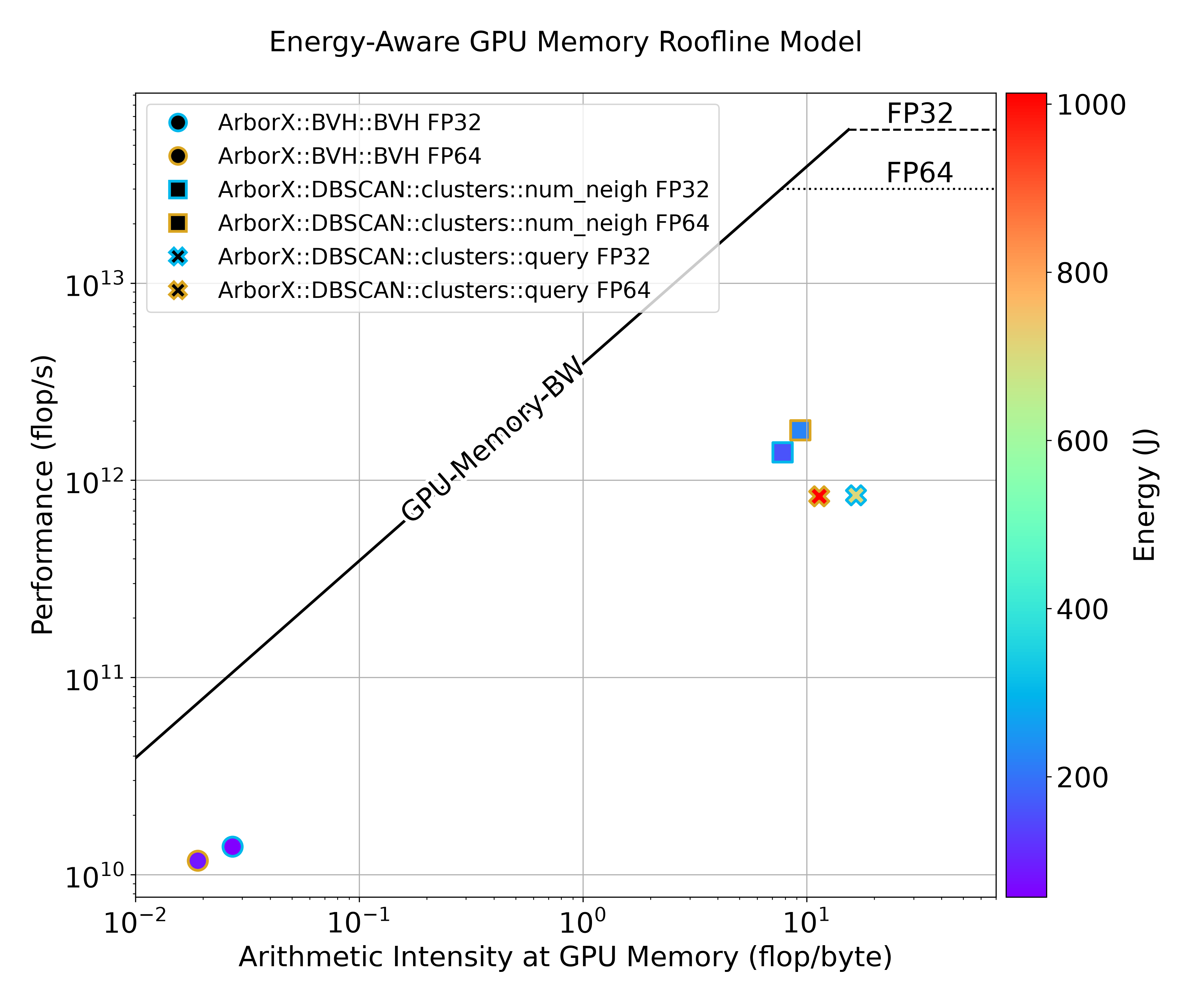}
    \caption{ArborX - Energy-Aware Roofline Model Large Workload. Outline color denotes precision (blue for FP32 and yellow for FP64); fill color denotes energy according to the color bar; marker shape denotes the kernel/region.}
    \label{fig:arborx_roof_lg}
\end{figure}

Figure \ref{fig:arborx_roof_lg} shows the Energy-Aware DRAM roofline model of the ArborX kernels for the large workload; \textit{DBSCAN::postprocess} is excluded due to lack of floating-point operations. Arithmetic intensity decreases from FP32 to FP64 by approximately 30\% in the \textit{BVH::BVH} and \textit{DBSCAN::clusters::query} kernels, while energy consumption increases by more than 40\%. In contrast, for \textit{DBSCAN::clusters::num\_neigh}, arithmetic intensity and performance throughput increase from FP32 to FP64 by 19.8\% and 29.6\%, respectively, accompanied by a 39.5\% increase in energy consumption.

Regarding the kernels' execution characteristics,  \textit{DBSCAN::clusters::query} and \textit{DBSCAN::clusters::num\_neigh} remain far from the FP32 and FP64 compute roofs. \textit{BVH::BVH} is influenced more strongly by memory bandwidth, similar to the LOBPCG kernels, while the other two kernels occupy the transition region between the memory-bandwidth and compute roofs.
 
For the small workload, arithmetic intensity follows a similar trend to that observed for the large workload. 
Compared with FP32, the performance throughput of the DBSCAN kernels increases by up to 57.7\% (\textit{DBSCAN::clusters::num\_neigh}) in FP64. The number of floating-point operations differs between FP32 and FP64 because the \textit{sqrt} function appears to produce an uneven increase in floating-point operations in our environment, thereby affecting the reported throughput.
Differences between precisions in energy consumption are minor. 
With respect to their proximity to the GPU memory-bandwidth roof, the kernels lie farther below the roof for the small workload than for the large workload, particularly \textit{BVH::BVH} and \textit{DBSCAN::clusters::num\_neigh}.
  
\begin{figure}[htbp]
    \centering
    \includegraphics[trim=22 22 48 82,clip, width=\linewidth]{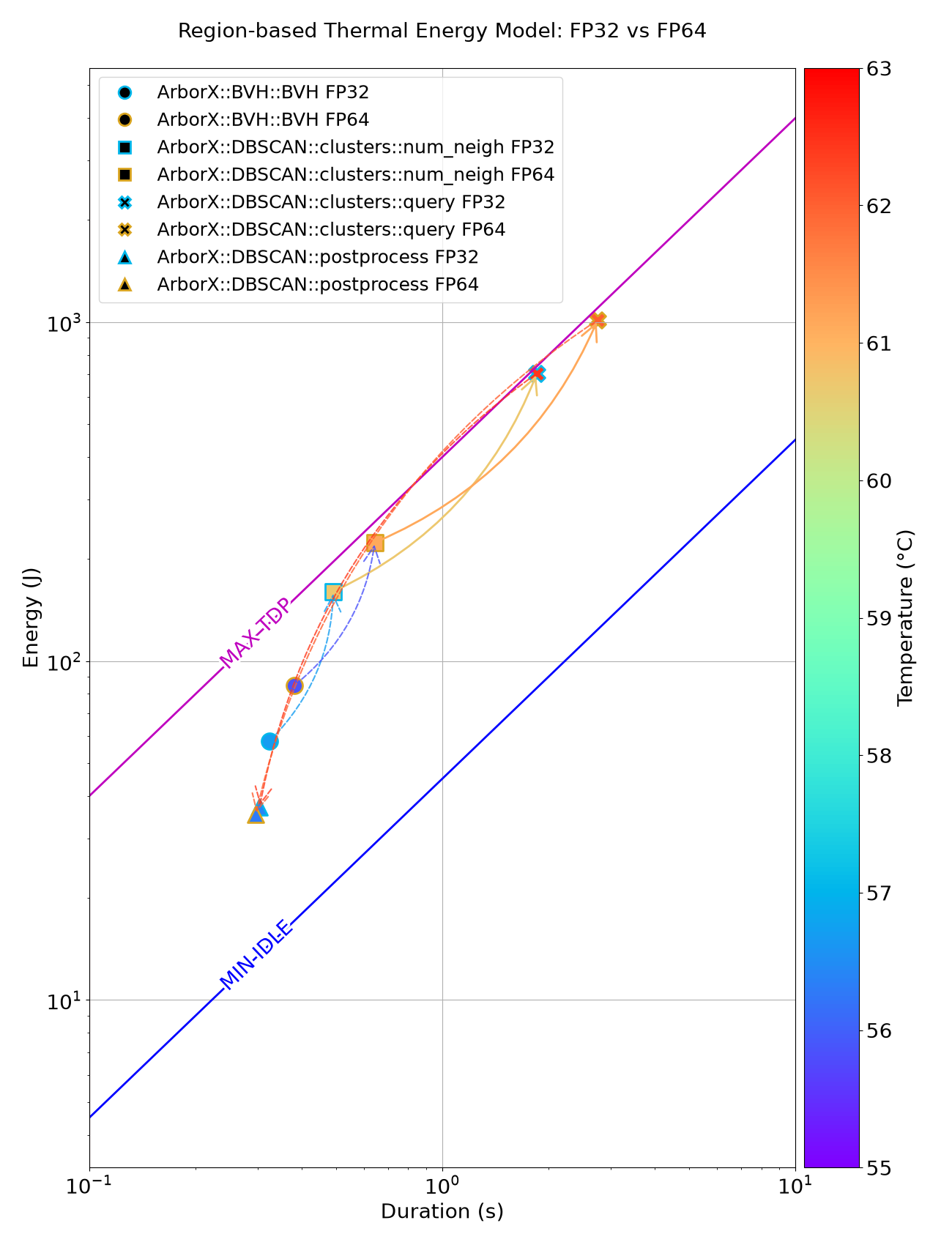}
    \caption{
    ArborX - Thermal-Aware Inter-Kernel Energy Model Large Workload. Outline color denotes precision (blue for FP32 and yellow for FP64); fill color denotes temperature according to the color bar; marker shape denotes the kernel/region; solid and dashed arrows indicate consecutive and nonconsecutive execution order, not heat transfer; diagonal lines are power references.
    }
    \label{fig:arborx_therm_lg}
\end{figure}
%large load
% ArborX::BVH::BVH FP32	0.32	57.94	56.71
% ArborX::BVH::BVH FP64	0.38	84.60	55.86
% ArborX::DBSCAN::clusters::num_neigh FP32	0.49	160.25	60.71
% ArborX::DBSCAN::clusters::num_neigh FP64	0.64	223.56	61.14
% ArborX::DBSCAN::clusters::query FP32	1.85	705.85	62.43
% ArborX::DBSCAN::clusters::query FP64	2.75	1012.64	62.00
% ArborX::DBSCAN::postprocess FP32	0.30	37.05	56.43
% ArborX::DBSCAN::postprocess FP64	0.30	35.30	56.29

Figure~\ref{fig:arborx_therm_lg} displays the Thermal-Aware Inter-Kernel Energy model of the ArborX kernels for the large workload.
%addresses RQ2 by retaining region costs and sampled thermal context in one view. 
The \textit{DBSCAN::clusters::num\_neigh} and \textit{DBSCAN::clusters::query} regions have the longest durations and largest energy costs among the four selected ArborX regions. They execute consecutively (solid directed lines) and lie near the TDP reference line. Their mean temperatures range from 60.7 to 62.4$^\circ$C, compared with 55.9 to 56.7$^\circ$C for \textit{BVH::BVH} and \textit{DBSCAN::postprocess}. The consecutive kernels are hotter and more energy-intensive than the other two. This suggests that neighboring regions can exhibit similar measured energy and temperature behavior, potentially forming thermal phases during application execution.
Across all four regions, the absolute temperature difference between the corresponding FP32 and FP64 measurements remains below 0.9$^\circ$C.

Temperature readings alone would miss the impact of precision on energy use. For \textit{DBSCAN::clusters::query}, FP32 consumes $705.9$~J at a temperature of $62.4$\,$^\circ$C, whereas FP64 consumes $1{,}012.6$~J at $62.0$\,$^\circ$C (means across seven runs). The energy sample standard deviation is small relative to the mean for both precisions. Thus, FP64 consumes 43.5\% more energy despite a slightly lower mean temperature reading. All seven FP64 temperature samples were 62$^\circ$C. The zero sample standard deviation does not establish identical underlying temperatures. Large-workload \textit{DBSCAN::postprocess} has slightly lower mean duration and energy in FP64, with energy differences small relative to the FP64 sample variation. Its lack of floating-point operations motivates checking the surrounding execution context, but does not identify the cause of the difference.

%small load
% ArborX::BVH::BVH FP32	0.198	22.302	54.571
% ArborX::BVH::BVH FP64	0.205	22.453	53.857
% ArborX::DBSCAN::clusters::num_neigh FP32	0.137	23.994	54.714
% ArborX::DBSCAN::clusters::num_neigh FP64	0.150	24.271	54.000
% ArborX::DBSCAN::clusters::query FP64	0.114	25.940	55.571
% ArborX::DBSCAN::postprocess FP32	0.111	12.823	57.143
% ArborX::DBSCAN::postprocess FP64	0.118	16.972	54.857

For the small workload, FP64 \textit{BVH::BVH} and \textit{DBSCAN::clusters::num\_neigh} take 3.5-9.3\% longer and consume up to 1.2\% more energy than FP32. \textit{DBSCAN::postprocess} provides a different comparison: FP64 has a lower mean temperature reading ($54.9$\,$^\circ$C versus $57.1$\,$^\circ$C), yet consumes 32.3\% more mean energy. Its mean energy is $17.0$~J in FP64 and $12.8$~J in FP32. 
However, the kernel has no floating-point operations. 
%We hypothesize that the large difference in energy consumption can be due to the previous kernels in the sequence which do contain floating-point operations.
Additionally, FP64 energy consumption exhibits greater variability, particularly for the short \textit{DBSCAN::clusters::num\_neigh} and \textit{DBSCAN::postprocess} kernels, while the FP32 and FP64 energy ranges overlap for all kernels.
One possible explanation for the energy variation of the short kernels is counter-update timing. The approximately 100~ms update interval~\cite{mlenergy_profiling_2026} is close to the 109--130~ms FP64 \textit{DBSCAN::postprocess} durations. Small timing shifts can cause a region's energy reading to include energy from preceding work or miss energy reported after the region ends. In contrast, the energy measured over the longer enclosing \textit{DBSCAN::total} region remains stable, suggesting that the variation in the postprocessing measurements is an artifact of sensor granularity rather than a variation in the overall energy consumption. 
%Thus, the Thermal-Aware Inter-Kernel Energy Model that provides execution-order context can help interpret region-level measurements. 
Temperature standard deviations are small in both precisions. 
We exclude paired energy comparisons for small-workload \textit{DBSCAN::clusters::query} because all seven FP32 durations fall below the 101~ms threshold.

The LOBPCG benchmark supports energy--duration comparisons, but not the model's application execution-order interpretation. Initial results show that all kernels operate close to the maximum TDP power, indicating GPU power-intensive behavior. Compared to FP32 and FP32/FP64, FP64 kernels generally have longer execution times, consume more energy, and exhibit comparable, though slightly lower, temperatures (up to 4.1\% lower). 
%Its temperatures describe repeated kernel execution. 
For $M=4{,}000{,}000$ (small workload), \textit{BLAS::gemm} with FP64 storage and arithmetic averages $70.3$\,$^\circ$C, versus $73.3$\,$^\circ$C with FP32 storage and FP64 arithmetic. Temperature standard deviations are small in both configurations. The 3.0$^\circ$C decrease above is the largest across the compared precision modes and two workloads. 

\subsection{Discussion}

%LOBPCG
Regarding LOBPCG, FP32 and mixed FP32/FP64 kernels generally exhibit comparable arithmetic intensity, while mixed precision delivers modestly lower performance throughput. FP32/FP64 mixed precision consumes more energy in most cases, although some kernels exhibit comparable or lower energy consumption. In contrast, the FP64 kernels consistently achieve lower arithmetic intensity and performance throughput while consuming more energy than both FP32 and FP32/FP64. These memory-bandwidth-limited kernels have different measured energy costs across precision modes. The bandwidth bottleneck alone therefore does not determine which precision minimizes energy consumption, revealing the need for incorporating energy consumption into
the Roofline analysis (RQ1).
FP32 provides the most favorable measured performance–energy trade-off.
However, in practice, the mixed-precision FP32/FP64 option may be favored as it enhances precision at a low additional cost compared to the single-precision FP32 option.
This is, for example, the case for Cholesky-QR~\cite{Yamazaki2015}.
Our findings therefore indicate that selecting FP32 or FP32/FP64 instead of FP64 can provide meaningful computational performance and energy benefits for LOBPCG with either workload (RQ3).

%ArborX
For ArborX, the effects of precision depend on the kernel region and workload. For the large workload, FP64 kernels generally run longer and consume more energy. For the small workload, runtime differences between precisions are generally smaller, while energy differences vary by kernel, with the FP32 and FP64 energy ranges overlapping for all evaluated kernels. 
Within the \textit{DBSCAN} region, \textit{DBSCAN::clusters::query} accounts for 69.0\% of the energy consumption in FP32 and 72.1\% in FP64.
Our ArborX findings indicate that selecting FP32 can reduce energy consumption and execution time, particularly for larger workloads, whereas this trend is less pronounced for smaller workloads (RQ3).
%there is large stdv for small load that overlaps the values between fp32 and fp64. by removinf the outliers the energy becomes similar between f32 and fp64, but fo temperaure is the only one that does not overlap!

The Thermal-Aware Inter-Kernel Energy Model reveals region energy costs, including costs absent from the floating-point Roofline, together with execution order and temperature. Temperature adds descriptive information, but the present results do not demonstrate an independent optimization benefit from that information. The practical implication is to prioritize duration and energy, then use the Roofline to investigate computational behavior. A subsequent thermal-control study would need to separate profiling effects from execution history and measure the response to a control action. The related work in Section~\ref{sec:thermal_related} motivates such studies. Within these limitations, our measurements can reveal neighboring regions with similar temperature and energy behavior, indicating potential thermal phases (RQ2).

Finally, setting flags to switch precision for the same kernel does not necessarily guarantee that the same number of floating-point operations (flop) will be executed. As observed in our measurements, the executed flop count can differ across precision levels, which may influence arithmetic intensity and overall performance. These differences should therefore be acknowledged when interpreting changes in the measured metrics across numerical precisions (RQ3).

\section{Conclusion and Future Work}
\label{sec:conclusion}

We presented two multidimensional performance visualization models—the Energy-Aware Roofline Model and the Thermal-Aware Inter-Kernel Energy Model—for analyzing Kokkos-based, GPU-accelerated HPC applications. We characterized the execution patterns of the primary non-overlapping kernels in LOBPCG and ArborX across numerical precisions and workloads.

The Energy-Aware Roofline Model connects computational kernel behavior to measured GPU energy. The Thermal-Aware Inter-Kernel Energy Model compares region duration and energy while retaining execution order and temperature, including regions without floating-point operations. 
The case of ArborX \textit{DBSCAN::clusters::query}, large workload, illustrates why these quantities must remain distinct: substantially different energy costs accompany similar temperature readings. Its dominant share of measured region energy establishes an optimization priority independently of temperature.

Compared with ArborX, LOBPCG exhibits more consistent trends across workloads, with larger and more systematic differences between FP32 and FP64 in arithmetic intensity, performance throughput, and energy consumption.
The dedicated LOBPCG benchmark characterizes repeated execution, while ArborX instrumentation retains application-region order. Both approaches sample GPU temperatures that can depend on execution history and instrumentation.

Future work will extend the application coverage and evaluate whether the identified regions offer useful control points for DVFS, power capping, or thermal scheduling. Such evaluation requires controlled thermal conditions and measurements of the intervention's effects on computational performance, GPU energy consumption, and temperature.

\section*{Acknowledgment}
\unless\ifblind
Research sponsored by the Laboratory Directed Research and Development Program of Oak Ridge National Laboratory, managed by UT-Battelle, LLC, for the U.S. Department of Energy.
This research used resources of the Experimental Computing Laboratory (ExCL) and the Oak Ridge Leadership Computing Facility (OLCF) at the Oak Ridge National Laboratory, 
which is supported by the Advanced Scientific Computing Research programs in the Office of Science of the U.S. Department of Energy under Contract No. DE-AC05-00OR22725.
\fi

Use of AI tools: ChatGPT-5.5 assisted with editorial revisions to the manuscript and with prototyping measurement scripts. All experiments were conducted on a physical testbed, and the authors verified all scientific claims and experimental results.
\bibliographystyle{IEEEtran}
\bibliography{references}{}

\end{document}